\documentclass[mti,article,accept,pdftex,moreauthors]{Definitions/mdpi} 
\firstpage{1} 
\pubvolume{1}
\issuenum{1}
\articlenumber{0}
\pubyear{2026}
\copyrightyear{2026}
\externaleditor{Firstname Lastname} 
\datereceived{22 January 2026 } 
\daterevised{19 February 2026 } 
\dateaccepted{ } 
\datepublished{ } 

\usepackage{booktabs}
\usepackage{tabularx}
\usepackage{changepage} 
\usepackage{array}
\usepackage{gensymb}

\usepackage[labelformat=simple]{subfig}

\DeclareCaptionLabelFormat{subcaptionlabel}{\normalfont(\textbf{#2}\normalfont)}
\newcommand{\blue}[1]{\textcolor{black}{#1}}

\usepackage[normalem]{ulem} 

\Title{Behavioral Engagement in VR-Based Sign Language \linebreak  Learning: Visual Attention as a Predictor of Performance and Temporal Dynamics}

\Author{Davide 
 Traini $^{1}$\orcidA{}, José Manuel Alcalde-Llergo $^{2}$\orcidB{}, Mariana Buenestado-Fernández 
 $^{3}$\orcidC{}, Domenico Ursino $^{1}$\orcidD{} and Enrique Yeguas-Bolívar $^{2, }$*\orcidE{}}

\address{%
$^{1}$ \quad Department of Information Engineering, Università delle Marche, Ancona, 60121, Italy
; d.traini@pm.univpm.it (D.T.); d.ursino@univpm.it (D.U.)\\
$^{2}$ \quad Department of Computer Science and AI, 
 Universidad de Córdoba, Córdoba, 14071, Spain; jmalcalde@uco.es\\
$^{3}$ \quad Department of Education, Universidad de Córdoba, Córdoba, 14071, Spain; marianabf@uco.es}

\corres{Correspondence: eyeguas@uco.es}

\abstract{Understanding how learners engage with immersive sign language training environments is essential for advancing virtual reality-based education and inclusion. This study analyzes behavioral engagement in SONAR, a virtual reality application designed for sign language training and validation. We focus on three automatically derived engagement indicators (Visual Attention (VA), Video Replay Frequency (VRF), and Post-Playback Viewing Time (PPVT)) and examine their relationship with learning performance in a sample of 117 university students. {\color{black} Participants completed a self-paced Training phase with 12 sign language instructional videos, followed by a Validation quiz assessing retention. We employed Pearson correlation analysis to examine the relationships between engagement indicators and quiz performance, followed by binomial Generalized Linear Model (GLM) regression to assess their joint predictive contributions. Additionally, we conducted temporal analysis by aggregating moment-to-moment VA traces across all learners to characterize engagement dynamics during the learning session.} Results show that VA exhibits a strong positive correlation with quiz performance (r = 0.76),
followed by PPVT (r = 0.66), whereas VRF shows no meaningful association. A binomial GLM confirms that VA and PPVT are significant predictors of learning success, jointly explaining a substantial proportion of performance variance ($pseudo-R^2$ = 0.83). Going beyond outcome-oriented analysis, we characterize temporal engagement patterns by aggregating moment-to-moment VA traces across all learners. The temporal profile reveals distinct attention peaks aligned with informationally dense segments of both training and validation videos, as well as phase-specific engagement dynamics, including initial acclimatization, oscillatory attention cycles during learning, and pronounced attentional peaks during assessment. Together, these findings highlight the central role of sustained and strategically allocated visual attention in VR-based sign language learning and demonstrate the value of behavioral trace data for understanding and predicting learner engagement in immersive environments.}

\keyword{virtual reality; sign language education; behavioral engagement; visual \linebreak  attention; learning analytics; {\color{black}head-gaze-tracking}; VR learning environments; immersive learning} 

\begin{document}

\section{Introduction}
\label{sec:introduction}

Virtual reality (VR) has demonstrated significant potential for enhancing learning outcomes across diverse educational domains~\cite{yu2022meta}. The~immersive nature of VR enables embodied, multimodal learning experiences particularly suited for skill-based domains such as sign language instruction~\cite{wang2024impact}, where learners must simultaneously encode handshape, orientation, movement, location, and~facial expressions. Recent systematic reviews confirm that VR-based sign language learning systems produce measurable improvements in learning outcomes and engagement compared to conventional instruction~\cite{berrezueta2025virtual}.

Despite these promising developments, a~critical gap persists in the literature: while studies document the general effectiveness of VR sign language systems, they have not systematically examined how objective behavioral engagement indicators extracted from learner interactions relate to user engagement. Contemporary VR learning research emphasizes behavioral metrics derived from built-in tracking: head and hand movements, movement entropy, and~exploration patterns. These metrics can be used as synchronous indicators of engagement and intrinsic motivation, providing real-time insight into cognitive \textls[-25]{states~\cite{mallek2024review}. Indeed, these behavioral indicators are scalable, continuously available from the learning platform, and~can support real-time adaptive feedback without additional~instrumentation.}

{\color{black} The present study addresses this gap by proposing the following research questions: \linebreak  (i)
} Which specific behavioral indicators, automatically captured during VR interaction, predict learning success in sign language training? and~(ii) Do these indicators evolve dynamically throughout the learning process? Addressing these questions is essential for developing adaptive VR learning systems that can provide real-time support based on learner engagement patterns. 
Specifically, we examined how three indicators of behavioral engagement, namely Visual Attention (VA), Post-Playback Viewing Time (PPVT), and~Video Replay Frequency (VRF), correlate with learning performance, and~we employed a Generalized Linear Model (GLM) regression to evaluate their combined predictive power. Finally, we analyzed how VA changes over the course of the interaction with the VR sign language application.

Our framework is grounded in established learning science theories: eye-tracking research establishing gaze duration as a proxy for cognitive engagement~\cite{AbeysingheEtAl2025,AdAdMo23}, cognitive load theory identifying fixation patterns as indicators of information processing~\cite{EkinEtAl2025}, and~self-regulated learning theory showing that voluntary content revisits reflect metacognitive strategies~\cite{WangEtAl2022,GalbraithEtAl2004}. Specifically, this study examines learner engagement in SONAR (the 
 interested reader can find the application at the following link: \url{https://sidequestvr.com/app/37864/} (accessed on 1 March 2026) 
), a~VR application developed within the ISENSE project for sign language training. This application
places users in the role of a hearing student interacting with a deaf avatar on a simulated university campus during the first day of class. The~participants, 117 university students, completed a self-paced Training phase that allowed video replays, followed by a Validation quiz aimed at assessing retention. We then evaluate whether the engagement indicators are associated with learning success by relating them to quiz performance through correlational analyses and regression-based modeling. In~addition, we move beyond aggregate measures by examining how visual attention changes over time throughout the interaction, highlighting temporal engagement patterns across the learning and assessment~phases.

Our principal findings are threefold: (i) VA and PPVT are strong, statistically significant predictors of learning performance, with~VA emerging as the dominant indicator; (ii) VRF shows negligible correlation with performance, suggesting that engagement quality supersedes repetition quantity; (iii) temporal analysis reveals structured engagement patterns with characteristic peaks during training and assessment phases, confirming the task-driven nature of the observed attention patterns. These findings extend recent work on learning analytics into sign language VR contexts~\cite{tao2025learning}. Practically, these metrics can be automatically computed to enable real-time attention-aware learning analytics, supporting adaptive instructional interventions such as refocusing prompts or targeted content~review.

The remainder of this paper is organized as follows. Section~\ref{sec:Related Work} reviews literature on VR for education, VR-based sign language instruction, and~engagement measurement in immersive environments. Section~\ref{sec:App Description} describes the SONAR VR application.  Section~\ref{sec:Research Questions and Concept of Engagement} formally introduces the engagement framework and research questions. Section~\ref{sec:Experimental Setting} presents the study design and data collection procedures. Section~\ref{sec:Results} reports quantitative analyses addressing the research questions. Section~\ref{sec:Discussion} interprets findings in the context of prior work and discusses limitations. Section~\ref{sec:Conclusions} summarizes contributions and implications for adaptive VR learning~systems.

\section{Related~Work}
\label{sec:Related Work}

This section positions our work in relation to existing research on VR-based education, inclusion, and~engagement measurement, with~a specific focus on sign language learning. Specifically, in~Section~\ref{subsec:Education and Virtual Reality}, we review research on VR for education, highlighting evidence on learning gains, motivational benefits, and~design challenges. Section~\ref{subsec:Virtual Reality and Inclusion} examines VR and inclusion, summarizing studies on accessibility, inclusive design, and~applications for deaf and hard-of-hearing (DHH) learners and sign language education. Finally, Section~\ref{subsec:Engagement in Virtual Reality} discusses engagement in VR, covering multidimensional models of engagement, behavioral and gaze-based indicators, and~the limited existing work on engagement metrics in VR sign language~learning.

\subsection{Education and~VR}
\label{subsec:Education and Virtual Reality}

{\color{black} 
Research consistently demonstrates VR's positive impact on learning outcomes, with~robust effects in healthcare~\cite{yu2022meta} and significant advantages over less immersive methods in scientific domains like biology and physics~\cite{hamilton2021immersive}. This effectiveness is largely attributed to VR's ability to facilitate embodied, constructivist learning experiences where users actively construct knowledge through direct interaction~\cite{mallek2024review}.

In language learning, VR instruction yields significant benefits for speaking, listening, and~vocabulary acquisition~\cite{chen2022effects, huang2021systematic}. Educational outcomes are further enhanced by optimizing learner–environment interactivity~\cite{harris2022assessing} and integrating gamification, which substantially boosts motivation and engagement compared to non-gamified approaches~\cite{yigitbas2024gamification}.

Despite these benefits, adoption is limited by barriers such as hardware costs, content scarcity, integration challenges, and~potential cybersickness~\cite{upadhyay2024barriers}. Consequently, hybrid approaches combining VR with conventional instruction often yield superior outcomes, suggesting VR is most effective as a complementary tool within broader instructional designs rather than as a standalone replacement~\cite{liu2023effects}.
}

\subsection{VR and~Inclusion}
\label{subsec:Virtual Reality and Inclusion}

{\color{black} 
VR advances inclusive education for students with disabilities, with~systematic reviews confirming enhanced social inclusion and skill development~\cite{odunga2025exploring,chalkiadakis2024impact}. Accessibility features such as text-to-speech, customizable avatars, and~high-contrast visuals significantly improve motivation and learning outcomes~\cite{moreno2025reinforcing}, particularly for intellectual disabilities where immersive VR outperforms non-immersive methods~\cite{franze2024immersive}. However, barriers persist, requiring collaborative user-centered design processes~\cite{moreno2025agile, creed2024inclusive}.

For DHH individuals and sign language learners, VR offers unique benefits through sign language support and real-time captioning~\cite{gonzalez2025inclusive}, guided by XR design principles emphasizing communication clarity~\cite{ubur2024dc}. Key innovations include AI gesture recognition, gamification, and~interactive environments~\cite{berrezueta2025virtual}, with~3D VR demonstrating superior engagement and efficiency over 2D systems~\cite{wang2024impact,el2022virtual}. Recent technical solutions, such as inverse kinematics tracking~\cite{immanuel2025accessible}, enhanced hand reconstruction~\cite{wen2024enhancing}, and~ISENSE sign detection frameworks~\cite{bisio2023training}, further expand VR's potential for accessible sign language education.
}

\subsection{Engagement in~VR}
\label{subsec:Engagement in Virtual Reality}

Engagement in educational settings is conceptualized as a multidimensional construct encompassing behavioral, emotional, and~cognitive dimensions~\cite{WangDegol2014,FredricksMcColskey2012}. In~VR learning environments, behavioral engagement reflects observable participation and interaction with the digital environment, cognitive engagement relates to mental effort and comprehension, and~emotional engagement captures motivational responses~\cite{LinEtAl2024}. VR environments are effective at eliciting engagement through immersion, interactivity, and~embodied learning experiences. A~systematic review of VR impact on student engagement found that VR positively influences cognitive engagement by facilitating understanding of complex and abstract knowledge through immersive learning experiences, behavioral engagement through increased active participation and motivation, and~affective engagement through emotional involvement and empathy development~\cite{LinEtAl2024}.

\subsubsection{Measuring Engagement in~VR}
\label{subsec:Measuring Engagement in VR}

A critical challenge in VR learning research is the operationalization and measurement of engagement. Recent work emphasizes the importance of objective, automatically-derived behavioral indicators that can be extracted from VR interaction logs without additional hardware or manual analysis~\cite{DuEtAl2025,BetetaEtAl2022}. These behavioral traces offer distinct advantages over physiological measures alone, as~they directly reflect learners' observable choices and the temporal allocation of effort during the learning task. Learning analytics provides a strong framework for analyzing students' learning processes and engagement via trace data collected from learning environments~\cite{WinterEtAl2024}.

\paragraph{Visual Attention and Gaze-Based Indicators}

{\color{black} 
Eye-tracking and head-tracking provide direct measures of behavioral engagement, utilizing metrics such as fixation duration and saccade frequency to assess cognitive attention allocation~\cite{AbeysingheEtAl2025,AdAdMo23,BozkirEtAl2023}. Research distinguishes between ambient and focal attention modes, noting that fixation duration correlates with task difficulty and cognitive load~\cite{NegiEtAl2020,EkinEtAl2025}. In~immersive environments, these gaze patterns serve as temporal indicators of engagement, where metrics like the ambient/focal coefficient track how learners process visual stimuli~\cite{AbeysingheEtAl2025}. For~skill-based tasks like sign language, sustained visual focus on the model is critical for encoding motor details, making gaze orientation a valid proxy for task-relevant engagement~\cite{bagher2021move}.
}

\paragraph{Behavioral Traces from Video Interaction Logs}

{\color{black} 
Automated interaction logs capture self-regulated learning strategies, where behaviors like pausing for reflection or selective re-watching reflect active engagement~\cite{WangEtAl2022,mo2022video}. These strategic adjustments allow learners to align content pacing with their processing capacity~\cite{GalbraithEtAl2004}, distinguishing high-performing active learners from passive viewers~\cite{YoonEtAl2021}. Machine learning analyses of such clickstream data have effectively modeled engagement states and predicted learning outcomes~\cite{MirriahiEtAl2017}, representing distinct facets of engagement such as effort and content interest~\cite{WinterEtAl2024}.
}

\paragraph{Multimodal Measurement Approaches}

{\color{black} 
While physiological measures can assess emotional states, behavioral metrics are more scalable and have been shown to correlate more strongly with learning outcomes~\cite{ZhangEtAl2024}. Reviews confirm the value of combining gaze metrics with behavioral data to assess cognitive states~\cite{GorinEtAl2024}. Consequently, this study concentrates on these objective, automatically derivable indicators to examine their relationship with performance in a motor-intensive sign language task.
}

\subsubsection{Motor Learning and Engagement in~VR}

{\color{black} 
Research on motor learning in immersive VR demonstrates its capacity to support the acquisition of complex movements under controlled conditions~\cite{HaarEtAl2021}. Notably, studies report that VR practice can yield performance improvements comparable to real-world training, establishing a functional link between visual attention patterns (e.g., quiet-eye) and motor skill acquisition~\cite{HarrisEtAl2020}.

Embodied learning perspectives further emphasize the sensorimotor coupling between perception, action, and~attention, where distinct interaction techniques directly influence performance~\cite{bagher2021move}. Recent work argues that learners refine internal motor representations through an integrated loop of observation and enactment~\cite{Lehrman2025EmbodiedIVR,Dastmalchi2024EmbodiedVR}. Consequently, sign language training in VR works as observational motor learning, where behavioral indicators, such as visual orientation toward the signer, serve as operational proxies for the attentional resources allocated to encoding complex manual and facial movements.
}

\subsubsection{Engagement in VR-Based Sign Language~Education}

As discussed in Section~\ref{subsec:Education and Virtual Reality}, VR systems for sign language education have demonstrated positive impacts on learning outcomes compared to traditional instructional materials~\cite{el2022virtual}, indeed gamified 3D VR environments produce higher engagement and retention than 2D alternatives~\cite{wang2024impact}. A~recent systematic review on VR in sign language education identifies gesture recognition and real-time feedback as key mechanisms through which VR sustains learner engagement~\cite{berrezueta2025virtual}. 

However, research directly linking objective behavioral engagement indicators extracted from VR logs to measurable learning outcomes in sign language education remains limited. Existing studies on VR for sign language education have primarily focused on general effectiveness assessments and usability evaluations~\cite{el2022virtual,wang2024impact,berrezueta2025virtual}. To~the best of our knowledge, no prior study has examined how automatically-derived behavioral engagement indicators from VR interaction correlate with learning performance in sign language training applications. This represents a significant gap in the literature. Indeed, while general research on VR learning demonstrates that behavioral traces can predict engagement and performance~\cite{DuEtAl2025,BetetaEtAl2022,ZhangEtAl2024}, the~specific application of such approaches to sign language learning remains unexplored. Moreover, the~broader question of how learners' engagement unfolds and changes over time during sign language VR interaction, has not been addressed in any prior~work.

The present study directly addresses both of these gaps: (1) by examining the relationship between automatically-captured behavioral engagement indicators and learning outcomes in a VR sign language context, and~(2) by analyzing how learners' engagement patterns evolve and change during interaction, revealing temporal dynamics of user engagement with the sign language VR application. This dual focus connects outcome-oriented analysis (how engagement relates to performance) with process-oriented analysis (how engagement changes over time), providing insights into both the effectiveness of behavioral metrics for predicting sign language learning and the dynamics of user engagement in immersive educational~environments.

\section{App~Description}
\label{sec:App Description}

\blue{The SONAR application is a VR platform developed within and supported by the ISENSE project (\url{https://www.isenseproject.eu/}  (accessed on 1 March 2026)), an~Erasmus+ funded initiative focused on inclusive higher education for deaf and hard-of-hearing students. The~application was designed and implemented by some members of the current authorial team as part of this project.}
The interested reader can find the application at the following link: \url{https://sidequestvr.com/app/37864/}  (accessed on 1 March 2026). It places users in the role of a hearing student interacting with a deaf avatar on a simulated university campus during the first day of class. The~experience is structured into two distinct phases: Training and Validation. Users can interact with the virtual environment using either controllers or hand~gestures.

\subsection{Overview}

The participant finds themselves on their first day of class at university, inside a tiered lecture hall. At~the beginning of the experience, they are able to navigate through the classroom and observe the different virtual elements present in the environment, such as the blackboard, classmates interacting with each other, textbooks, and~other objects. In~one of the seats, Essie, a~deaf classmate, can be found, with~whom the participant can interact to communicate using sign~language.

Upon approaching Essie, the~application presents the user with two options: Training 
 and Validation. Once one of these options is selected, whose differences will be explained later, a~conversation in sign language is initiated. During~this interaction, the~participant is shown how to sign several basic phrases that can be used to introduce oneself and establish an initial contact with a classmate. The~conversation consists of 12 interactions, each composed of one user contribution followed by a response from the deaf classmate’s avatar. These interactions are listed in Table~\ref{tab:dialogue_interactions}.

\begin{table}[H]

\caption{Interactions between the hearing participant and the deaf avatar during the VR~experience.}
\label{tab:dialogue_interactions}

\begin{tabular}{p{6.5cm}p{6.5cm}}
\toprule
\textbf{Hearing Participant} & \textbf{Essie (Deaf Avatar)} \\
\midrule
1. Can you read my lips? & Yes. \\
2. Hi! & Hi! How are you? \\
3. Is this seat available? & Yes, please, take it. \\
4. What's your name? & My name is Essie. \\
5. Do you have a name sign?,\\ What's your name sign? & My sign is (ISENSE sign). \\
6. Do you live here? & Yes. \\
7. I'm glad to meet you. & Me too. \\
8. Is it your first year? & Yes. \\
9. You like Information Technology (IT),\\don´t you? & More or less. \\
10. Have you already met any other students? & No. \\
11. Don´t worry.\\I can introduce you to many other friends. & That’s great. \\
12. Here comes the teacher.\\The lesson is about to start. & All right, then. \\
\bottomrule
\end{tabular}
\end{table}

The SONAR application supports six languages: English, Spanish, Italian, German, Austrian, and~Polish. Communication is facilitated through two modes: animated avatars that display International Sign Language (for English) or the corresponding national sign languages for the other languages, and~prerecorded videos featuring different signed statements. These videos were recorded by professional sign language interpreters in the respective sign languages of each~country.

\subsection{\color{black} Training and Validation~Phases}

As previously described, the~experience is structured into two distinct phases: a Training phase, designed to support sign language learning, and~a Validation phase, intended to assess the user’s acquired signing~skills.


When the Training phase is selected, the~conversation takes place directly, guiding the user through the signing of several basic phrases. During~this phase, the~user is allowed to practice the signs associated with each phrase while observing the natural flow of the dialog. All signed expressions performed by both the sender (hearing participant) and the receiver (deaf avatar) can be replayed multiple times. {\color{black} In this way, learners can repeatedly observe the movements from within a first-person situated interaction with Essie rather than through a flat video.}

In the Validation phase, the~same conversational sequence used in Training is presented again. However, whenever the user is required to act as the sender, they must choose one out of three possible signing options displayed in the interface. {\color{black} Each option is presented as a short video embedded in the virtual classroom, and~the user selects their answer directly in the 3D scene (by pointing the controller or using a simple confirm gesture), maintaining the feeling of interacting with Essie in a shared space rather than clicking on a 2D menu}. A~scoring mechanism is employed for assessment: a correct selection on the first attempt yields 50 points, 25 points are awarded if the correct option is selected on the second attempt, and~0 points are assigned if both attempts are~incorrect.

{\color{black} Methodologically, the~Training phase focuses on information encoding through self-regulated interaction, allowing learners to control the pace and focus of their learning. Conversely, the~Validation phase is designed for information retrieval and assessment, as~it tests the learner's ability to recognize and reproduce the signs in a contextualized 3D~environment.}

\begin{figure}[H]
    
    \includegraphics[width=0.95\linewidth]{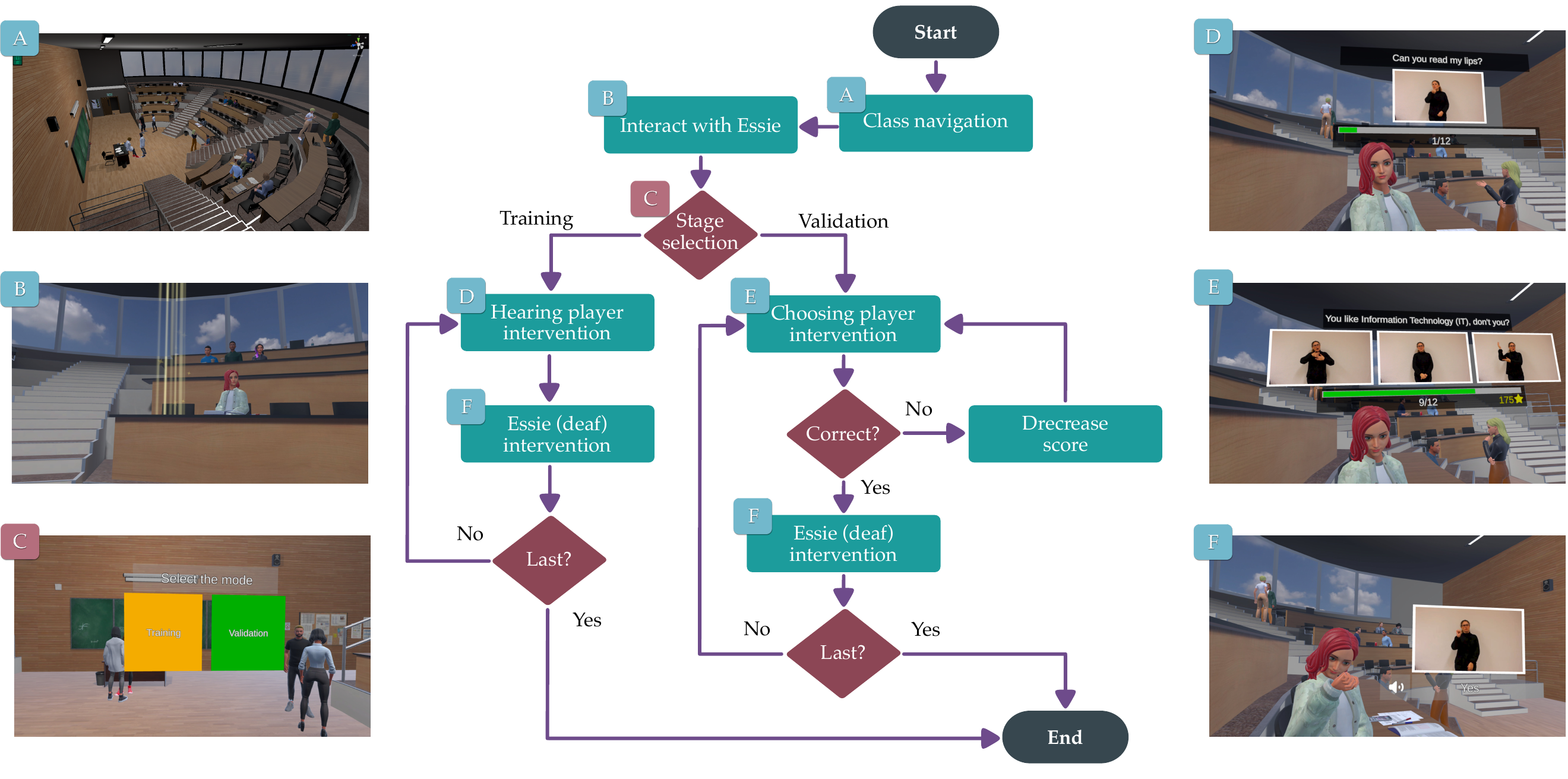}
    \caption{\hl{Flow} 
 diagram of the SONAR~experience.}

    \label{fig:flow}
\end{figure}

Figure~\ref{fig:flow} illustrates the flow diagram of the proposed experience, detailing the sequence of interactions and decision points that structure the {\color{black}learner’s} progression. The~diagram summarizes the main stages of the experience, from~initial classroom navigation and interaction with the deaf avatar to the selection between Training and Validation modes. It also highlights the different interaction dynamics, feedback mechanisms, and~finish conditions that control each phase, providing an overview of the system logic underlying the sign language learning and assessment~process.

\unskip

\subsection{Deployment Setup and Interaction with the~Environment}
\label{subsec:deployment_setup}

Users can interact with the virtual environment through Meta Quest devices, supporting six degrees of freedom (6DoF) tracking. Locomotion can be performed using the standard handheld controllers, enabling both linear displacement and teleportation-based navigation within the virtual~classroom.

To facilitate continuous sign language practice and reduce the need for physical controllers during interaction, a~controller-free locomotion mechanism based on hand gestures has also been implemented. In~this mode, users can navigate the environment by extending both arms. Additional interactions, such as selection or confirmation, are performed through intuitive hand gestures, including pointing with the index finger and confirming selections by closing the fist. {\color{black} This implementation allows for a complete interaction without interrupting signing activities and preserves the benefits of embodied, spatially situated observation that differentiate the VR experience from conventional 2D video-based materials. The~immersive 6DoF setup in SONAR enables learners to slightly change their viewpoint relative to Essie, thereby inspecting handshapes, orientations, and~trajectories from different angles while remaining within a coherent 3D classroom scene. The~choice to use a VR environment instead of a 2D video interface is consistent with prior work showing that immersive VR via HMDs can outperform less immersive methods in terms of learning outcomes and engagement~\cite{hamilton2021immersive,mallek2024review}}

\section{Research Questions and Concept of~Engagement}
\label{sec:Research Questions and Concept of Engagement}

This section clarifies the conceptual framework used to interpret engagement in the SONAR environment and states the study’s aims. In~Section \ref{subsec:Engagement in the SONAR Environment}, we defined the concept of engagement applied to our VR learning scenario, and~we introduced the behavioral indicators used to operationalize it: VA, VRF, and~PPVT. In~Section \ref{subsec:Research Questions}, we present the Research Questions that guide the~analysis.

\subsection{Engagement in the SONAR~Environment}
\label{subsec:Engagement in the SONAR Environment}

As discussed in Section~\ref{subsec:Engagement in Virtual Reality}, engagement in educational settings is commonly conceptualized as comprising cognitive, behavioral, and~affective dimensions~\cite{WangDegol2014,FredricksMcColskey2012}. In~VR learning environments, cognitive engagement is associated with learners' attention and mental effort toward mastering the instructional content; behavioral engagement reflects observable participation and interaction within the environment, and~affective engagement encompasses motivation, interest, and~emotional involvement. In~line with recent work on immersive learning and learning analytics, the~present study concentrates on behavioral manifestations of engagement that can be captured in a non-invasive way through the interaction between the learner and the VR platform~\cite{DuEtAl2025,BetetaEtAl2022}.

In the context of our SONAR application, three aspects of learner behavior are particularly informative for understanding engagement with the material. The~first aspect is how consistently learners direct their attention toward the signing model during instruction~\cite{AbeysingheEtAl2025,AdAdMo23}.
The second aspect is the extent to which learners actively decide to revisit the instructional material~\cite{WangEtAl2022,GalbraithEtAl2004}. The~third aspect is how long they continue to attend to the visual representation of the signs after playback has finished~\cite{NegiEtAl2020,WinterEtAl2024}. For~this reason, in~order to predict user engagement, we choose three behavioral variables: VA, VRF, and~PPVT.

VA 
 is defined as the proportion of time in which the user is oriented toward the video screen displaying the signing avatar during both Training and Validation Phases. This definition is grounded in recent research, which has established that gaze-based indicators are valid proxies for attention and cognitive engagement in VR and immersive video learning environments~\cite{AbeysingheEtAl2025,AdAdMo23}. Research has shown that fixation duration on task-relevant regions predicts both engagement and learning outcomes~\cite{NegiEtAl2020}. In~sign language learning, sustained visual focus on the signer is essential to perceive manual articulation, facial expressions, and~body movements that jointly convey linguistic information. Studies on embodied learning and sign language teaching with virtual avatars emphasize that learners must maintain stable attention on the signing space to accurately encode handshape, movement, and~location~\cite{bagher2021move,Dastmalchi2024EmbodiedVR}. This makes VA a natural behavioral correlate of task-relevant attention in sign language~contexts.

{\color{black} Operationally, we defined a rectangular Region of Interest (ROI) positioned in the center of the screen, subtending 10\degree{} of the visual field, corresponding to the central attention zone. For~each frame, we computed the percentage of the target video screen visible within this ROI. We chose head pose over direct eye-tracking because the used VR devices do not provide integrated eye-tracking capabilities. This design choice ensures that the behavioral metrics can be extracted from widely accessible consumer VR hardware, enhancing the scalability and reproducibility of engagement analytics in educational VR applications. Furthermore, this choice is supported by other works demonstrating that head pose effectively identifies attended regions in VR environments~\cite{higgins2022head}.}

{\color{black} VRF denotes the number of times each training video is reloaded by a learner during the Training phase, normalized by the total time spent in the Training phase. This normalization accounts for individual differences in session duration.} This indicator is motivated by research on self-regulated learning and behavioral engagement, which shows that voluntary decisions to revisit instructional content often reflect metacognitive strategies aimed at clarifying misunderstandings, reinforcing memory traces, or~preparing for subsequent assessment~\cite{WangEtAl2022}. Video analytics research has identified that self-regulated learning in video-based contexts involves strategic use of playback controls, including rewinding and replay~\cite{GalbraithEtAl2004, WangEtAl2022, mo2022video}. Research on learner clusters has identified that active engagement patterns, characterized by frequent interaction, predict better learning outcomes~\cite{YoonEtAl2021}.

In video-based and online learning contexts, fine-grained analyses of pausing, rewinding, and~replay sequences have been used to identify segments that learners find difficult and to characterize deeper engagement with challenging parts of the content~\cite{GalbraithEtAl2004,YoonEtAl2021}. Learners employ diverse self-regulated strategies with video playback controls, including, pausing for reflection, and~selective replaying of difficult segments~\cite{GalbraithEtAl2004}. In~the specific case of sign language training, replaying videos allows learners to repeatedly observe complex handshapes, orientations, and~movements, which can support the refinement of perceptual discrimination and the formation of robust mental representations of the signs~\cite{wang2024impact}. The~strategic, selective nature of replay indicates that learners are engaging in deliberate practice and self-monitoring~\cite{WangEtAl2022}. For~these reasons, VRF is interpreted as an indicator of strategic re-engagement with the instructional~material.

PPVT {\color{black} is a metric extracted from the analysis of the Training phase}; it captures the duration for which a video remains visible on the screen after its active playback has finished, until~the learner either replays it or initiates the next video. {\color{black} Operationally, during~this period, the~final frame of the video remains frozen and displayed on the virtual screen, while learners retain full freedom of movement within the VR environment; indeed, they can move their heads, move their hands to practice the sign, or~shift their body position. The~controlled nature of the SONAR classroom environment ensures minimal external distractions during this window; no other animated elements or interactive objects compete for attention.} This temporal window is interpreted {\color{black} as an operational proxy for} post-presentation processing. During~this interval, learners may mentally rehearse the observed signs, inspect articulatory details, or~reflect on how the signs relate to previously learned~items.

The use of a PPVT {\color{black} as an indirect behavioral indicator} is grounded in eye-tracking research showing that dwell time (i.e, the~duration of sustained attention on task-relevant stimuli) is associated with deeper cognitive processing~\cite{NegiEtAl2020}. In~multimedia learning environments, longer viewing durations on relevant stimuli have been associated with improved retention of information and deeper cognitive engagement~\cite{WinterEtAl2024}. More specifically, fixation duration research has demonstrated that learners engaged in active cognitive processing show specific patterns of fixation times that differ from passive viewing, and~that these patterns predict subsequent learning outcomes~\cite{NegiEtAl2020}. {\color{black} While PPVT does not directly measure internal cognitive states, it serves as a behavioral trace that may reflect the time allocated to such processes.}

Embodied learning and motor learning studies further show that extended observation following the demonstration of a movement can facilitate internal simulation and consolidation of motor representations~\cite{bagher2021move,Dastmalchi2024EmbodiedVR}. This is especially important for visually mediated skills such as sign language production, where learners must internally simulate and encode complex motor patterns observed in the signing model. The~presence of the video on-screen after playback completion provides an opportunity for this post-presentation cognitive processing and motor consolidation. Therefore, PPVT is interpreted as an {\color{black}  an operational indicator that may index} how much additional cognitive and perceptual processing learners devote to each signed phrase once the dynamic motion has~ended.

\subsection{Research~Questions}
\label{subsec:Research Questions}

The analyses in this work focus on the following research questions:

\begin{itemize}
    \item RQ1: How are behavioral indicators of engagement (namely VA, VRF, and~PPVT) related to learning performance in an immersive sign language training scenario?

    \item RQ2: How do learners' attention and engagement evolve while viewing the sign language videos, and~what general temporal patterns emerge in these indicators?
\end{itemize}

These research questions aim to clarify both the association between behavioral engagement metrics and learning performance, and~the temporal dynamics of engagement while learners interact with sign language videos in VR. Our first research question (RQ1) adopts an outcome-oriented perspective, examining whether automatically-derived behavioral indicators predict learning success. Prior research has shown that aggregate engagement measures computed across sessions correlate with learning outcomes in diverse educational contexts~\cite{DuEtAl2025,BetetaEtAl2022,WinterEtAl2024}, but~none of them are related to sign language~environment.

Our second research question (RQ2) adopts a process-oriented perspective, examining how learners' engagement changes over time during the interaction. Since analysis of behavioral traces reveals typical engagement trajectories and temporal patterns in how learners interact with learning content~\cite{PrakashEtAl2025,AdAdMo23}, by~aggregating temporal profiles of visual attention across learners, it is possible to identify general engagement patterns and reveal when engagement peaks or declines during learning activities~\cite{AdAdMo23,ShadievLi2023}.

Together, these two research questions integrate outcome-oriented and process-oriented perspectives, providing complementary insights into how engagement operates in VR-based sign language~learning.

\section{Experimental~Setting}
\label{sec:Experimental Setting}

The study involved 117 university students who attended a VR sign language training session lasting approximately 15--20 min, {\color{black} followed by a validation phase of approximately 10--15 min}. Each participant was exposed to the same set of interactions (videos; see Table~\ref{tab:dialogue_interactions}), which were always presented in a fixed order. {\color{black} The sample comprised participants with varying levels of prior VR exposure: 56.1\% had no previous experience with VR environments, while the remaining 43.9\% reported few prior VR interactions. Regarding sign language knowledge, 86\% of participants had no prior experience with sign language.}

{\color{black} As described in Section~\ref{subsec:deployment_setup}, the~application provides two different types of interaction: handled and controller-free. In~the present study, all participants used the controller-free hand-tracking modality exclusively. This methodological choice was made to maximize immersion for the sign language learning environment, where continuous hand availability is essential for learners to practice signs while observing instructional videos. The~controller-free modality also eliminates potential confounds from controller manipulation during behavioral data collection, ensuring that head orientation tracking for Visual Attention measurement and interaction timestamps for PPVT and VRF calculation reflect genuine learning behaviors rather than device-handling actions.
}

Each student completed two consecutive phases. In~the Training phase, participants freely watched the 12 videos one by one, with~the possibility to replay each video (once it had finished) as many times as desired. In~the subsequent Validation phase, they answered a quiz in which each item required selecting the video that matched a target signed phrase previously seen in the Training phase. Each quiz item in the Validation phase was associated with exactly one video from the Training phase, as~described in Section~\ref{sec:App Description}. \blue{A Training–Validation session comprising 12 items was implemented because this number allowed the complete protocol to be conducted within approximately 30 min, while maintaining an adequate number of observations per participant. Increasing the number of items would have extended the session duration and potentially introduced fatigue effects that could have confounded the performance measures.}

User interaction information were collected throughout the Training and Validation phases. {\color{black} Specifically, VA is calculated in both Training and Validation phases, while VRF and PPVT are only related to the Training Phase.} For RQ1, we first examined the distribution of quiz scores and then computed Pearson correlations between the three behavioral engagement variables (i.e., VA, VRF, PPVT) and the performance variable defined as the number of correct answers obtained in the Validation quiz (Section~\ref{subsubsec:correlation_engagement_performance}). We then complemented this analysis with a regression study in which the proportion of correct answers was modeled as a function of VA, PPVT, and~VRF in order to assess their joint and unique contributions to learning performance (Section~\ref{subsubsec:regression_engagement}). In~relation to RQ2, the~same time-stamped interaction logs were used to characterise the general evolution of learners' attention and engagement during video viewing, by~analysing the temporal patterns of VA, across the Training and Validation phase and within individual videos (Section~\ref{subsec:rq2}).

{\color{black} It is important to note that this study employs a purely correlational design without control conditions. All sessions were conducted under standardized conditions in controlled laboratory settings, using the Meta Quest default parameters; instructions and task structures were kept constant across participants, ensuring consistent exposure to the training materials and interaction procedures. This design allows us to identify predictive relationships but does not permit causal inferences regarding the effects of engagement on learning performance. Experimental validation through controlled designs would be necessary to establish causality.}

\section{Results}
\label{sec:Results}

In this section we show all the results associated to RQ1 and RQ2. Specifically, in \mbox{Section~\ref{subsec:rq1}}, we investigate which of the three behavioral indicators is the most informative for predicting the users' learning performance. In~Section~\ref{subsec:rq2}, we discuss the dynamic of VA during the Training and Validation phases, highlighting the peaks and the trend of the~curve.

\subsection{RQ1: How Are Behavioral Indicators of Engagement Related to Learning Performance in an Immersive Sign Language Scenario?}
\label{subsec:rq1}

This subsection presents quantitative analyses of user interaction data recorded during the VR sign language session, focusing on the association between engagement metrics and overall performance. In~Section~\ref{subsubsec:correlation_engagement_performance} we perform a correlation analysis between behavioral indicators and performance, while in Section~\ref{subsubsec:regression_engagement} we use a binomial GLM to predict the number of correct answers using the three behavioral indicators as~features.

\subsubsection{Correlation Between Engagement and~Performance}
\label{subsubsec:correlation_engagement_performance}

\begin{figure}[H]

\includegraphics[width=0.6\textwidth]{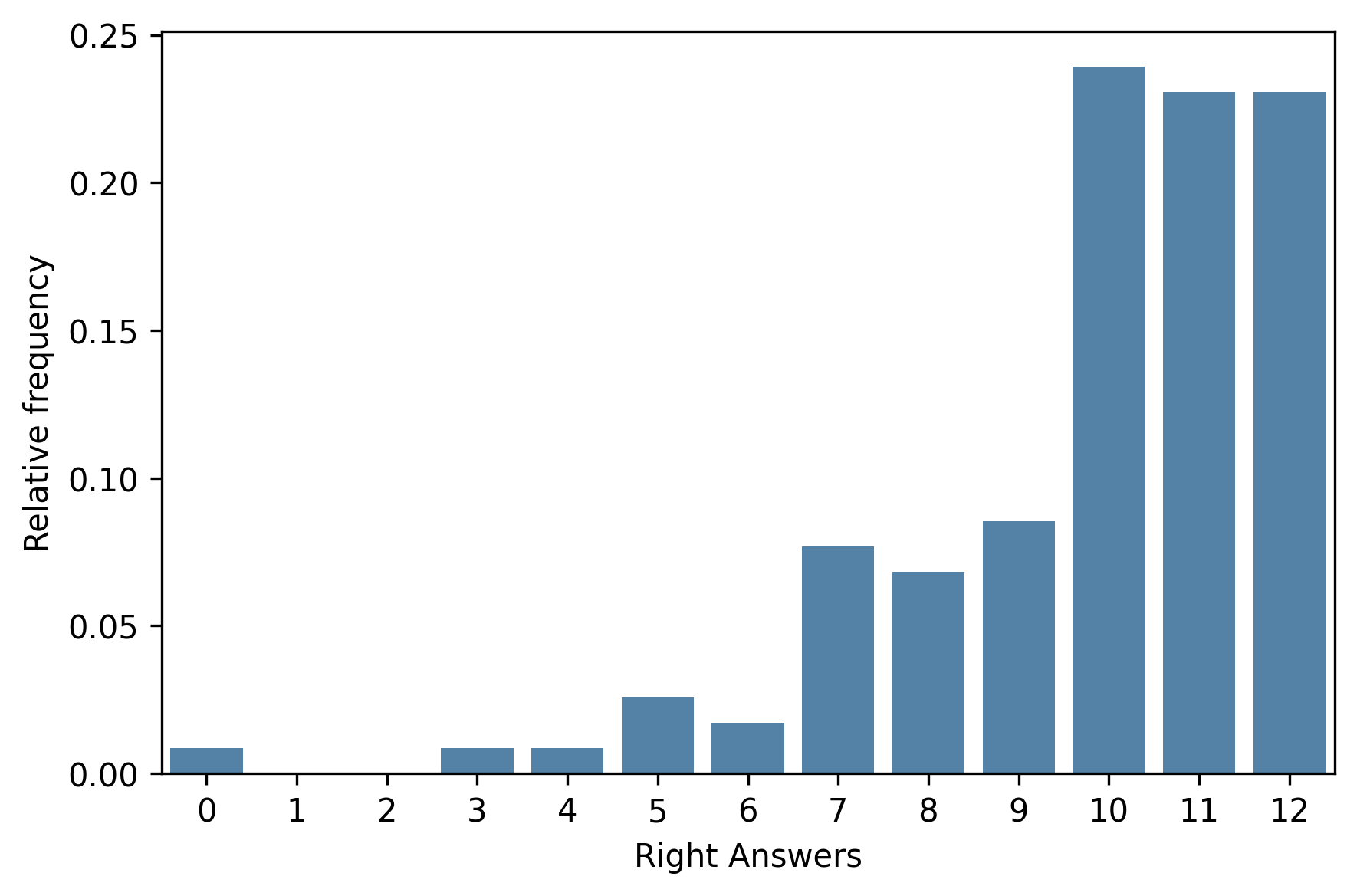}
\caption{Distribution of correct answers in the Validation~phase.}
\label{fig:grad_distr}
\end{figure}

In Figure~\ref{fig:grad_distr} we show the distribution of correct answers in the Validation phase. The~distribution is shifted toward higher scores, with~a mean of $\mu = 9.846$ and standard deviation of $\sigma = 2.172$. This indicates that most participants successfully answered the majority of quiz~items.

To investigate the relationship between behavioral engagement and learning outcomes, we computed Pearson correlation coefficients between the three engagement metrics defined in Section~\ref{sec:Research Questions and Concept of Engagement} and quiz~performance.

\begin{figure}[H]

\subfloat[PPVT vs quiz performance.]{
\includegraphics[width=0.30\textwidth]{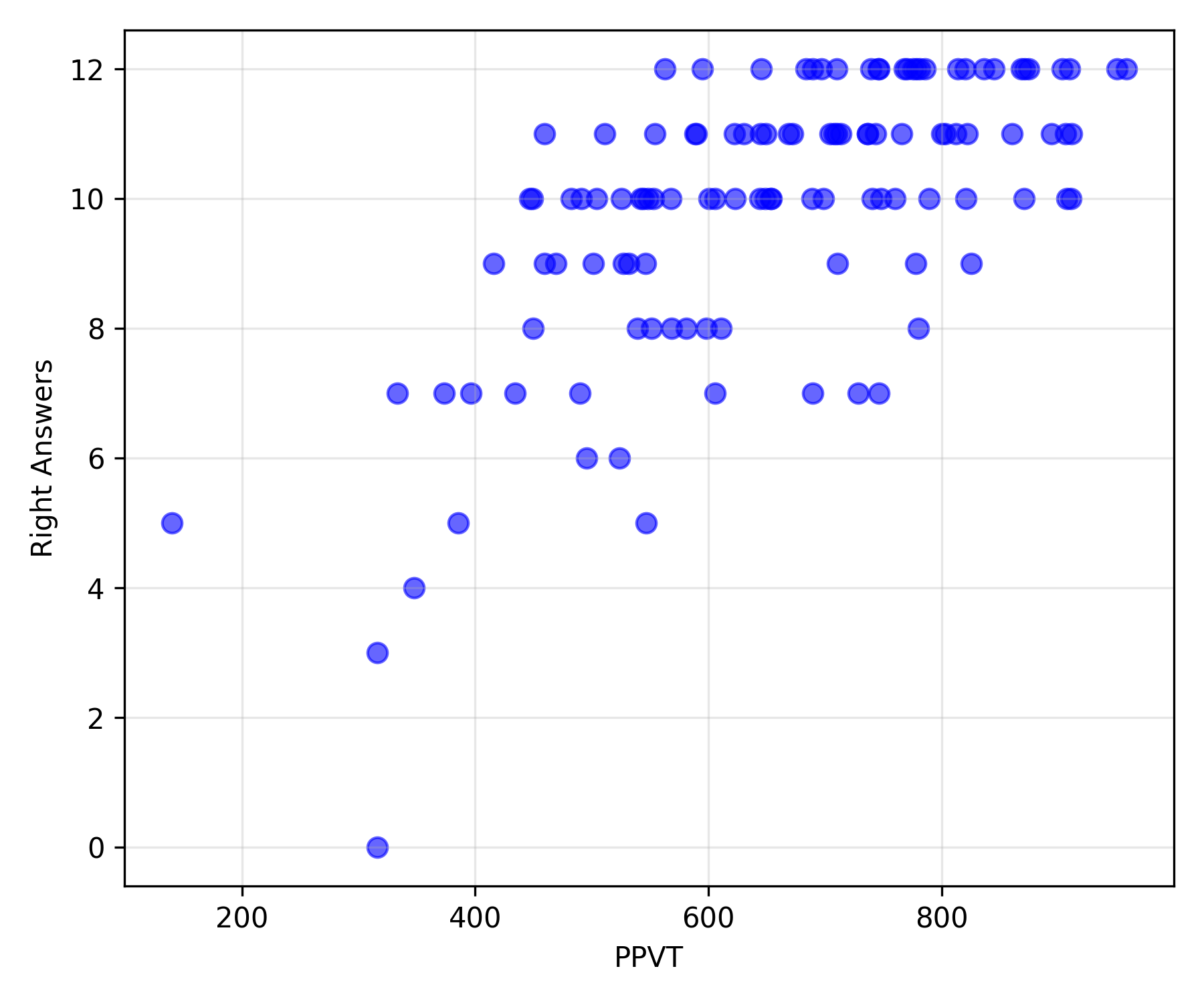}
}
\hfill
\subfloat[VA vs quiz performance.]{
\includegraphics[width=0.30\textwidth]{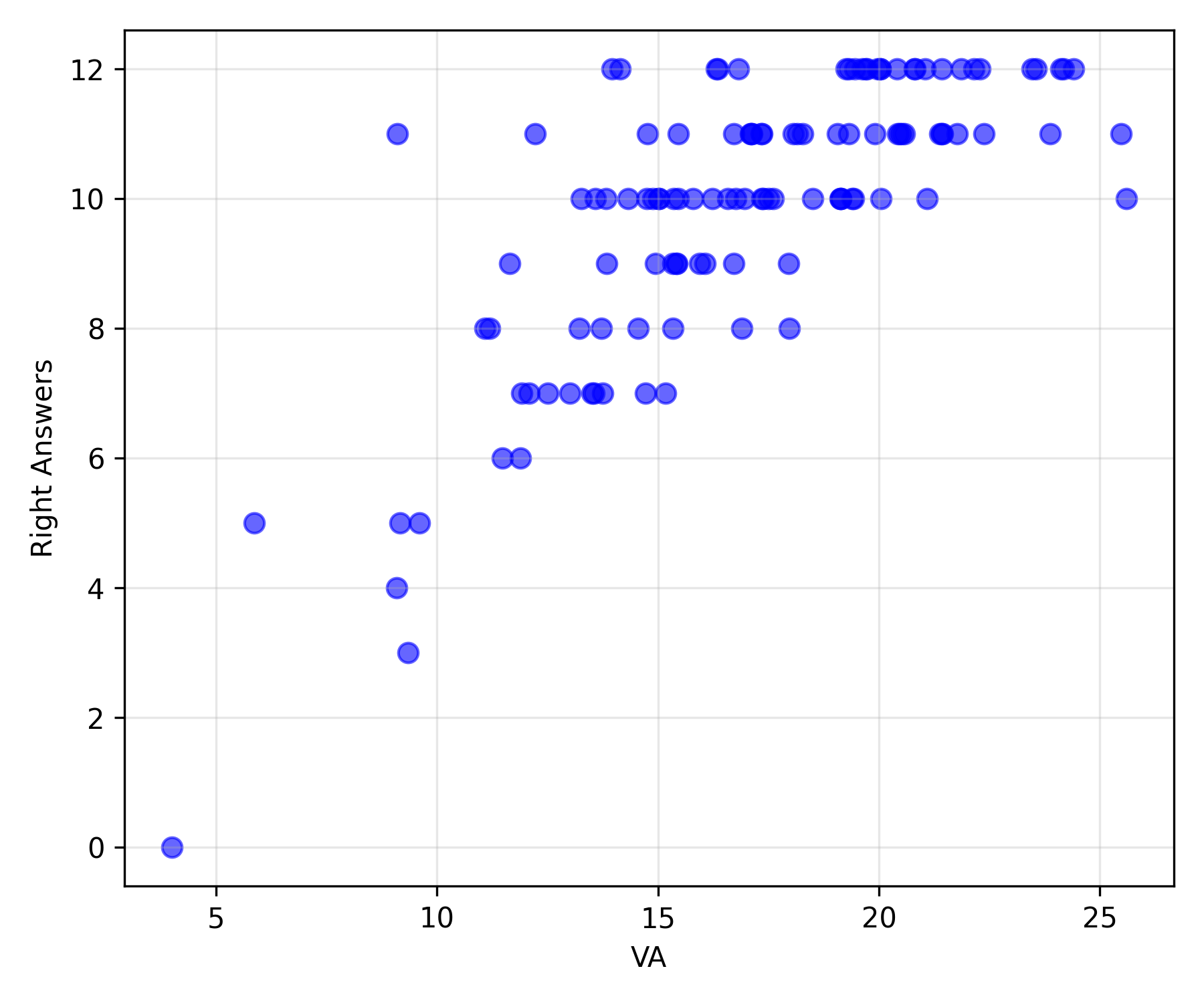}
}
\hfill
\subfloat[VRF vs quiz performance.]{
\includegraphics[width=0.30\textwidth]{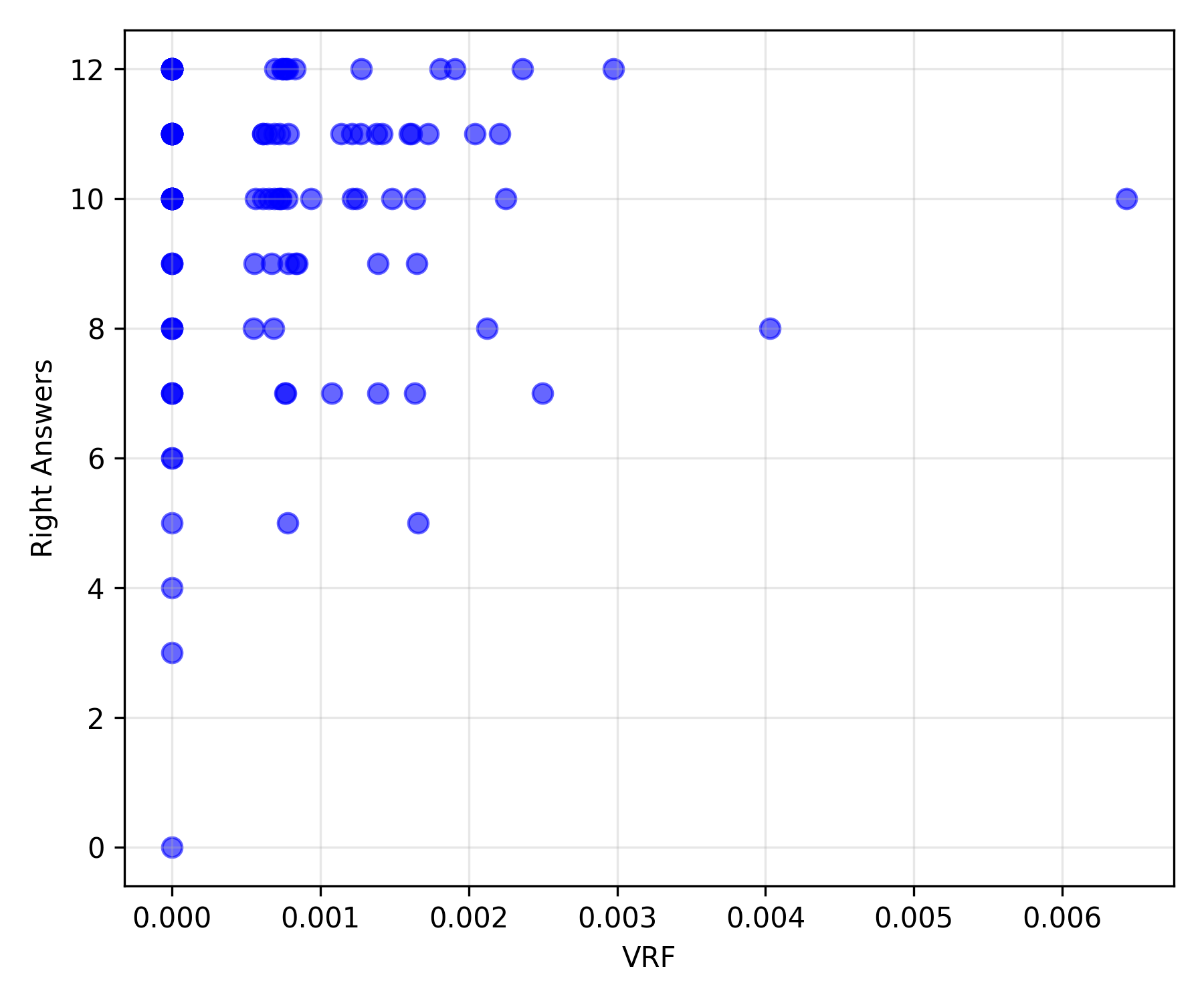}
}
\caption{Relationships 
 between quiz performance and the three behavioral engagement metrics: (\textbf{a})~Post-Playback Viewing Time (PPVT); (\textbf{b}) Visual Attention (VA); (\textbf{c}) Video Replay Frequency (VRF).}
\label{fig:corr_all}
\end{figure}

PPVT exhibited a strong positive correlation with performance ($r = 0.66$, $p < 0.001$), as~illustrated in Figure~\ref{fig:corr_all}a. Participants who maintained videos visible on the screen for extended durations after playback ended, ostensibly engaging in reflection or mental rehearsal, achieved notably better learning~outcomes.

\vspace{-9pt}

VA showed the strongest correlation with quiz performance ($r = 0.76$, $p < 0.001$), as~shown in Figure~\ref{fig:corr_all}b. This metric, operationalized as the proportion of time the user's gaze was oriented toward the video screen, reflects sustained attention directed toward instructional content. The~strong positive relationship indicates that participants who maintained consistent visual focus on the signing model achieved substantially higher quiz~scores.

In contrast, VRF displayed a negligible and non-significant correlation with performance {\color{black}($r = 0.02$, $p = 0.811$)}, as~shown in Figure~\ref{fig:corr_all}c. This finding indicates that the raw number of times participants replayed videos does not meaningfully predict learning success. The~lack of correlation suggests that engagement quality matters substantially more than the sheer quantity of~repetitions.

\subsubsection{Regression Analysis of Behavioral Engagement and~Performance}
\label{subsubsec:regression_engagement}

To examine the joint contribution of behavioral engagement metrics to learning performance, we modeled quiz performance as a function of the three engagement indicators. At~the participant level, performance was expressed as the proportion of correct answers in the Validation phase. We fitted a binomial GLM with logit link function, using standardized versions of Visual Attention ($\text{VA}_z$), Post-Playback Viewing Time ($\text{PPVT}_z$), and~Video Replay Frequency ($\text{VRF}_z$) as predictors 
:
\begin{equation}
\label{eq:glm}
\text{logit}(\text{prop\_correct}) = \beta_0 + \beta_1 \,\text{VA}_z + \beta_2 \,\text{PPVT}_z + \beta_3 \,\text{VRF}_z
\end{equation}

Table~\ref{tab:glm_results} presents the estimated coefficients on the log-odds scale, their standard errors, z-statistics, significance levels, and~corresponding Odds Ratios. $\text{VA}_z$ and $\text{PPVT}_z$ emerge as positive and highly significant predictors of performance. $\text{VRF}_z$ shows a near-zero and non-significant coefficient, indicating negligible additional predictive value once the other metrics are included. The~model achieves a Cox-Snell pseudo-$R^2$ of 0.8309, indicating that these engagement metrics jointly explain a substantial proportion of variance in learning outcomes. It is worth noting that this substantial predictive power is partly attributable to the specific distribution of the performance data (Figure \ref{fig:grad_distr}). Since the majority of participants achieved high scores (mean = 9.846), the~dataset presents a `ceiling effect' where instances of low performance are distinct exceptions. The~GLM effectively leverages the engagement metrics ($\text{VA}_z$ and $\text{PPVT}_z$) to discriminate these specific cases of lower performance from the high-performing majority, resulting in high model fit indices typical of datasets with this type of skewness. \textcolor{black}{For this reason, the~pseudo-$R^2$ value should not be interpreted as evidence of near-complete explanatory power in a general learning context, but~rather as an indication that, within~this restricted and skewed performance range, the~selected behavioral indicators separate high- and low-performing learners relatively~well.}

\begin{table}[H]

\caption{\color{black}Binomial GLM for the proportion of correct answers. Predictors are standardized ($z$-scores). Coefficients are on the log-odds scale with Standard Errors (SE), $z$-statistics, $p$-values, and~Odds Ratios (OR).}
\label{tab:glm_results}

\setlength{\tabcolsep}{5.6mm}  
\begin{tabular}{lrrrrr}
\toprule
\textbf{Predictor} & \textbf{Coefficient} & \textbf{SE} & \boldmath{$z$} & \boldmath{$p$}\textbf{-Value} & \textbf{OR} \\
\midrule
Intercept & 1.8194 & 0.086 & 21.104 & $<$0.001 & --- \\
$\text{VA}_z$ & 0.7558 & 0.094 & 8.040 & $<$0.001 & 2.129 \\
$\text{PPVT}_z$ & 0.4359 & 0.088 & 4.931 & $<$0.001 & 1.546 \\
$\text{VRF}_z$ & $-$0.0339 & 0.076 & $-$0.448 & 0.654 & 0.967 \\
\bottomrule
\end{tabular}
\end{table}

The positive coefficients for $\text{VA}_z$ and $\text{PPVT}_z$ indicate that, holding other variables constant, participants exhibiting higher VA and longer PPVT achieve better quiz performance. Since the predictors are standardized, a~one-unit increase corresponds to a one standard deviation increase in the respective engagement metric. Interpreted as Odds Ratios, a~one standard deviation increase in $\text{VA}_z$ multiplies the odds of correctly answering a quiz item by approximately {\color{black} 2.13}, while a one standard deviation increase in $\text{PPVT}_z$ multiplies the odds by approximately {\color{black} 1.54}. These effect sizes demonstrate that both sustained visual focus on the signing model and additional processing time after playback are strongly associated with successful recognition and comprehension of signed~phrases.

Conversely, the~coefficient for $\text{VRF}_z$ is near zero and not statistically significant {\color{black}(\mbox{$p = 0.654$})}. Its Odds Ratio is approximately {\color{black}0.97}, representing essentially no change. This result indicates that, once VA and PPVT are accounted for, the~VRF provides no meaningful additional predictive information. The~qualitative aspects of engagement appear far more informative for predicting learning outcomes than the quantity of repetitions~alone.

\begin{figure}[H]

\subfloat[]{
\includegraphics[width=0.30\textwidth]{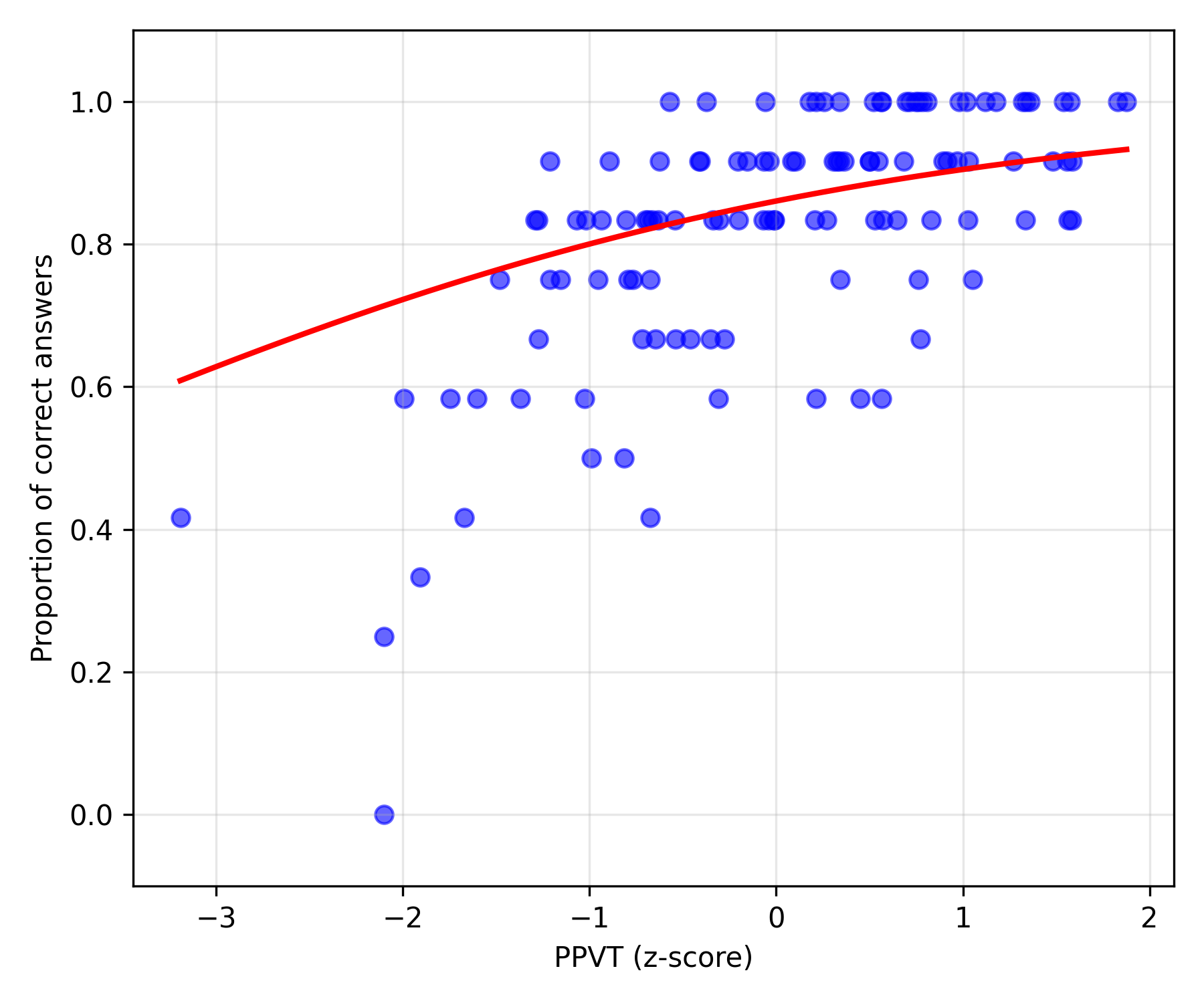}
\label{fig:regr_ppvt}
}
\hfill
\subfloat[]{
\includegraphics[width=0.3\textwidth]{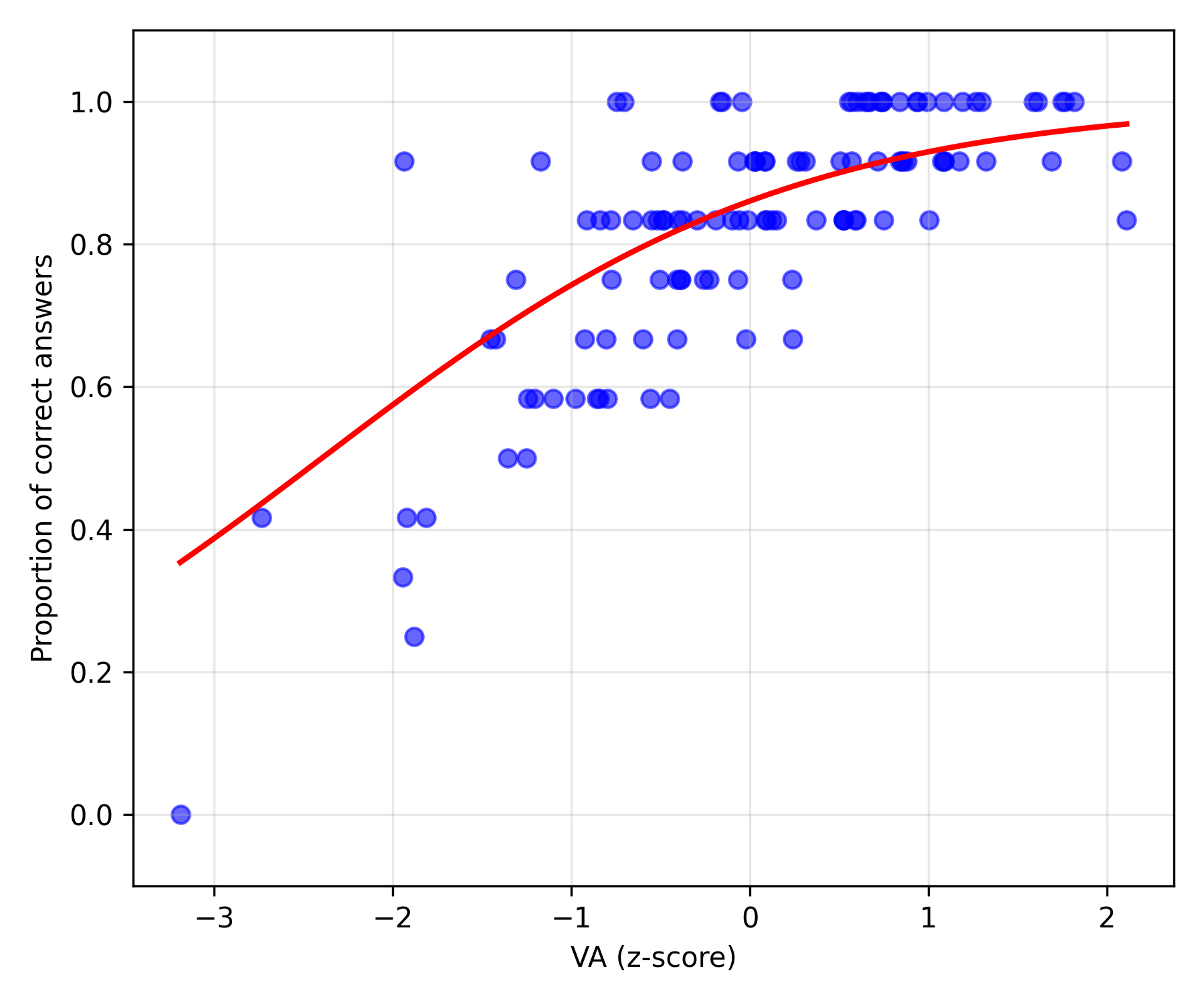}
\label{fig:regr_va}
}
\hfill
\subfloat[]{
\includegraphics[width=0.30\textwidth]{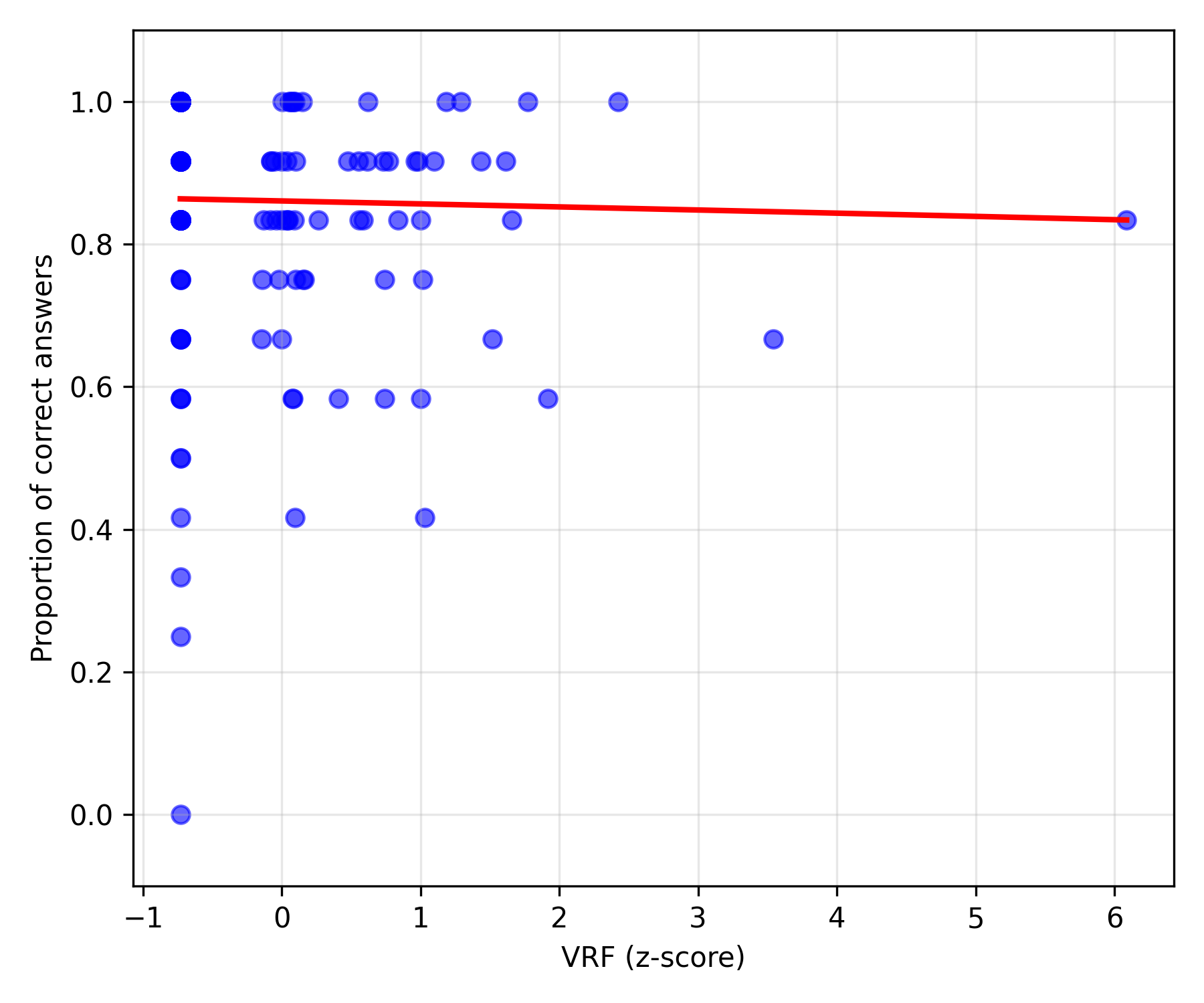}
\label{fig:regr_vrf}
}
\caption{Predicted 
 proportion of correct answers from the binomial GLM as a function of each standardized engagement metric. The red curve shows the model-predicted probability of a correct response when varying one predictor while holding the others at their mean~values: (\textbf{a}) Post-Playback Viewing Time ($\text{PPVT}_z$). Effect 
 of $\text{PPVT}_z$ on predicted quiz performance; (\textbf{b}) Visual Attention ($\text{VA}_z$). Effect of $\text{VA}_z$ on predicted quiz performance; (\textbf{c}) Video Replay Frequency ($\text{VRF}_z$). Effect of $\text{VRF}_z$ on predicted quiz performance.}
\label{fig:regr_all}
\end{figure}

Figure~\ref{fig:regr_all} displays the GLM predictions for the proportion of correct answers as a function of each standardized engagement metric. In~all three panels, blue dots represent observed participant quiz scores, while the red curve shows the model-predicted probability of a correct response when varying one predictor while holding the others at their mean~values.

Across the plots, $\text{PPVT}_z$ and $\text{VA}_z$ show clearly increasing regression curves, visually confirming that longer PPVT and higher VA are associated with higher predicted performance. The~nearly flat curve for $\text{VRF}_z$ confirms that changes in VRF have a negligible impact on predicted performance, consistent with its non-significant coefficient in Table~\ref{tab:glm_results}.

\vspace{-6pt}
\unskip

\subsection{RQ2: What Do These Behavioral Metrics Suggest About Learners' Engagement Processes?}
\label{subsec:rq2}

Given the findings from RQ1, we focus exclusively on VA as the primary indicator for characterizing engagement processes throughout the training activity. This choice is justified by both empirical and theoretical considerations. In~RQ1, VRF showed negligible correlation with learning performance and contributed no significant predictive value in the regression model. PPVT, while correlated with performance, represents a static summary measure aggregated across the entire training session and lacks temporal resolution. In~contrast, VA can be measured and sampled at fine temporal scales, providing moment-to-moment information about how learners allocate their visual focus during both Training and Validation~activities.

The decision to use VA as a time-varying indicator of engagement aligns with established research on gaze-based measures in learning environments. Eye-tracking research has consistently demonstrated that gaze direction and fixation patterns serve as direct, real-time indicators of where learners direct their attention and, by~extension, where they focus their cognitive resources~\cite{AbeysingheEtAl2025,AdAdMo23}. In~immersive learning environments, these gaze-based metrics have been operationalized as temporal indicators of engagement precisely because they can be sampled continuously throughout the learning process, allowing researchers to characterize how engagement unfolds and changes across time~\cite{AdAdMo23}. Advanced analytical frameworks have developed temporal profiles of VA, aggregating moment-by-moment gaze data across learners to derive general engagement patterns that reveal when engagement peaks or declines during learning activities~\cite{AbeysingheEtAl2025}.

Consequently, in~our analysis of RQ2, we use VA as the primary temporal indicator of engagement. Rather than treating engagement as a single static aggregate, we analyze how VA evolves across the sequence of Training and Validation videos, identifying temporal patterns and dynamics that characterize learners' engagement processes during sign language VR training. This process-oriented perspective complements the outcome-oriented analysis from RQ1, revealing not only whether engagement predicts performance, but~also how engagement unfolds dynamically as learners interact with the instructional~material.

\begin{figure}[H]

\includegraphics[width=\textwidth]{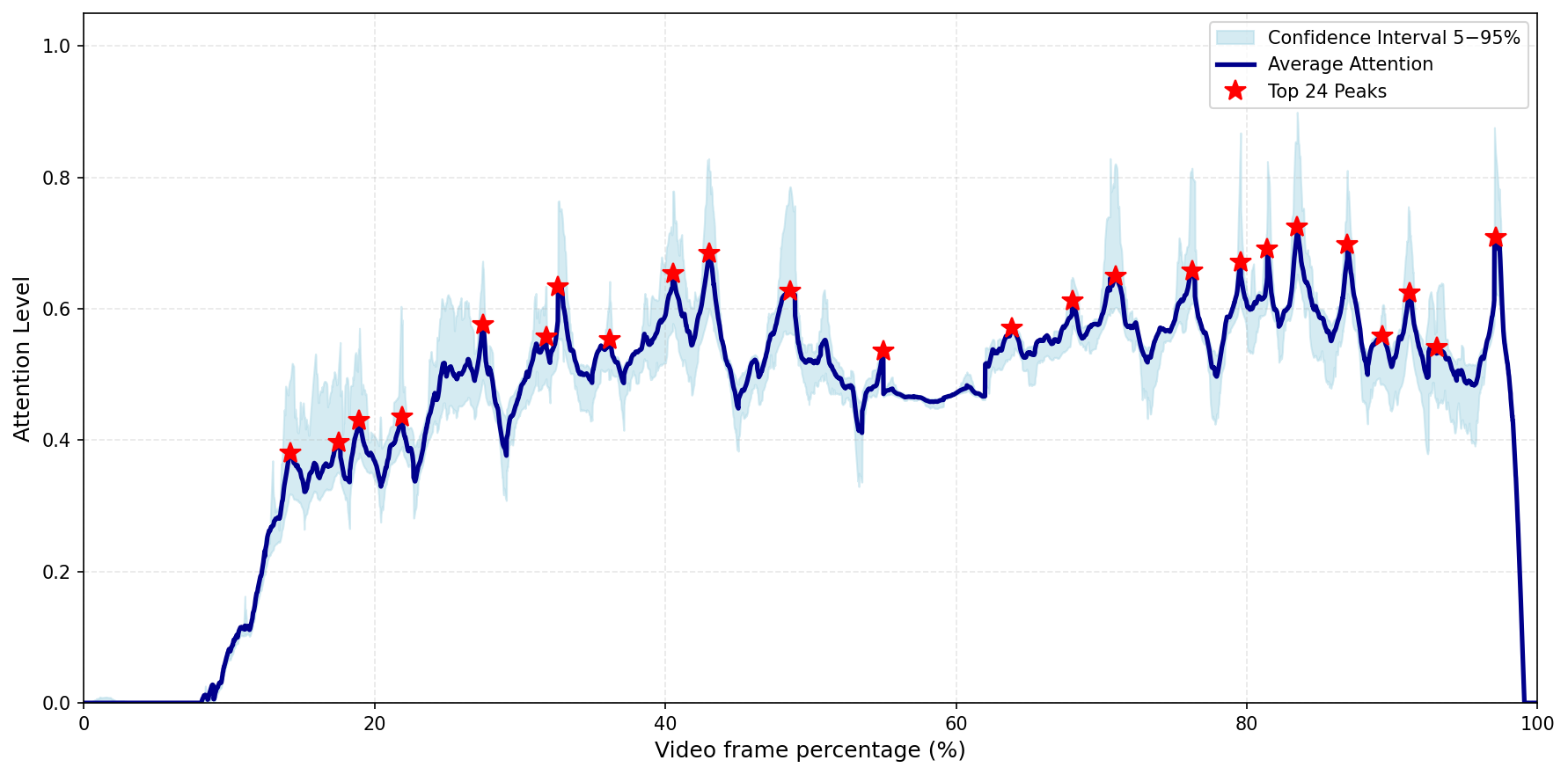}
\caption{Temporal 
 profile of aggregated VA across all participants. The~blue curve represents mean VA normalized across video frame progression, with~the light blue band indicating the 95\% confidence interval. Red stars mark the 24 highest peaks, corresponding to moments of particularly intense visual focus during video~playback.}
\label{fig:temporal_va}
\end{figure}

Figure~\ref{fig:temporal_va} presents the aggregated temporal profile of VA across all 117 participants throughout their complete interaction with the VR application. The~profile reveals distinct patterns of engagement that correspond to the different phases and cognitive demands of the learning~experience.

During the initial portion of the profile (0 to approximately 15 percent), VA increases gradually as learners begin their interaction with the application. This slow ramp-up reflects a characteristic pattern in immersive learning environments where participants require time to acclimate to the virtual environment and transition their attentional resources from environmental exploration to focused engagement with instructional content. Research on immersive learning has documented that learners initially allocate attention to environmental exploration and familiarization with interface elements before settling into focused engagement with the learning material~\cite{ShadievLi2023}.

Following this initial phase, the~profile shows 12 distinct peaks, each corresponding to one of the 12 training videos presented during the instructional phase. These peaks represent moments of heightened VA during the presentation of signing demonstrations, indicating that learners direct concentrated visual focus toward specific moments within each video. The~emergence of distinct peaks suggests that different moments within each training video elicit varying levels of visual attention. In~multimedia learning, learners modulate their attentional allocation based on perceived salience and task relevance of different content segments~\cite{NegiEtAl2020,AbeysingheEtAl2025}. In~sign language learning, these peaks likely correspond to moments when manual articulation, handshape transitions, or~facial expressions are most prominent or when linguistic information is~densest.

Between consecutive peaks, VA shows noticeable declines before recovering toward the next peak. This oscillating pattern within the training phase reflects characteristic attention dynamics in which learners maintain high focus during moments of high informational density, then experience transient reductions in visual attention in the intervals following these moments. One interpretation is that after observing key articulatory or linguistic information, learners engage in internal processing, reflection, or~mental rehearsal rather than maintaining continuous visual focus on the signing model. This pattern is consistent with cognitive science research on attention and working memory, which demonstrates that learners benefit from periods of distributed attention that allow for encoding and consolidation of perceived information~\cite{NegiEtAl2020}. The~declines may also reflect natural variations in motor behavior and saliency across the content, where moments of lower visual prominence naturally redirect learner~attention.

A marked transition occurs at approximately 55 to 63 percent of the profile, where VA drops to a lower, more stable baseline. This plateau corresponds to the transition from the training phase to the Validation phase. Following the plateau, the~profile exhibits a second set of 12 peaks, each corresponding to one of the 12 quiz items in the Validation phase. These peaks appear as learners are presented with recognition and decision-making tasks, indicating moments when quiz items capture their visual attention. The~appearance of these validation peaks indicates that learners return to the task with increased visual focus during assessment activities.. These validation peaks indicate sustained engaged visual processing during the recognition and selection tasks required by the~quiz.

Immediately following the completion of the validation phase (near 100 percent of the profile), VA shows a sharp, dramatic collapse. This precipitous decline reflects the moment at which learners exit the interactive application and conclude their engagement with the system. The~sharp drop indicates that the elevated VA observed throughout both phases is task-driven and stimulus-dependent, contingent on the presence of active instructional or assessment~activities.

The temporal structure of VA revealed in this aggregated profile illuminates how learners allocate visual attention across different types of cognitive engagement in sign language learning. The~distinct peaks corresponding to training videos reflect intensive visual perception when learners extract motor and linguistic information from signing demonstrations. The~presence of peaks during validation, suggests that learners maintain engaged visual processing when retrieving and recognizing previously observed signs. The~intervening periods of lower VA may represent cognitive processing windows during which learners consolidate perceived information or mentally rehearse observed~signs.

\section{Discussion}
\label{sec:Discussion}

This study provides empirical evidence regarding the role of behavioral engagement indicators in predicting learning performance in VR-based sign language training, addressing a significant gap in the literature. From~the analysis of the users' interaction, we extracted insights about the three behavioral indicators. From~a correlational perspective, VA emerges as the most correlated variable with respect to validation quiz performance ($r = 0.76$, $p < 0.001$), a~result aligned with established literature on eye-tracking and cognitive load theory~\cite{NegiEtAl2020}. Indeed, in~the context of sign language, maintaining stable visual attention on the model is essential for accurately perceiving manual articulation, facial expressions, and~body movements that jointly convey linguistic information~\cite{AbeysingheEtAl2025,AdAdMo23}. Our results extend these established theoretical principles to the specific domain of VR-based sign language education, a~context previously lacking  empirical~investigation.

PPVT emerges as the second most correlated variable ($r = 0.66$, $p < 0.001$), {\color{black} suggesting that the time spent observing the video after playback completion may serve as an operational proxy for post-presentation cognitive processes} critical for consolidating motor representations~\cite{Dastmalchi2024EmbodiedVR}. During~this temporal window, learners likely engage in internal simulation, perceptual review of observed signs, and~reflection on how learned elements relate to prior knowledge. This result {\color{black} is consistent with} hypotheses derived from motor learning research, which has demonstrated that extended observation of movement demonstrations, particularly after the movement concludes, facilitates internal simulation and consolidation of motor representations~\cite{bagher2021move}. {\color{black} However, our study did not directly measure internal simulation processes; the observed correlation between PPVT and performance suggests a relationship that may involve such mechanisms, but~alternative explanations (e.g., simple memory rehearsal, comparative analysis between signs, or~individual differences in processing speed) cannot be excluded based on our data alone.} In terms of cognitive load theory, post-playback time may represent a window in which learners deliberately reduce extraneous cognitive load (visual distraction from motion) while maintaining germane cognitive processing (internal review of learned material), {\color{black} although the specific cognitive mechanisms underlying this behavioral pattern warrant further investigation through complementary methods such as concurrent verbalization or neuroimaging.}

Notably, VRF {\color{black} ($r = 0.02$, $p = 0.811$)} shows negligible and non-significant correlation with performance, in~contrast with some of the related literature~\cite{GalbraithEtAl2004}. In~reality, this reflects a fundamental principle of motor learning research: the quality of deliberate practice outweighs the amount of exposure~\cite{NegiEtAl2020}. This result is conceptually interesting and invites reconsideration of simple models of self-regulated learning that equate replay frequency with engagement level or metacognitive sophistication. In~the context of visually mediated motor acquisition tasks such as sign language, attentional quality during initial viewing appears to outweigh the quantity of subsequent repetitions. An~alternative interpretation is that replay may sometimes represent a compensatory strategy employed by learners who failed to maintain focused visual attention during initial viewing; in this case, low replay would indicate effective initial perception rather than inferior engagement. This hypothesis merits empirical investigation in future studies employing experimental designs that deliberately manipulate replay~opportunities.

The binomial GLM indicates that both standardized VA and PPVT are significant predictors of quiz performance (regression coefficients, odds ratios). Increases in VA are associated with a higher likelihood of correct responses (odds ratio), while PPVT also shows a positive, though~more moderate, contribution (odds ratio). In~contrast, VRF does not exhibit a meaningful effect on performance (\textit{p}-value, odds ratio), suggesting a negligible association with quiz outcomes. The~model demonstrates strong explanatory power (Cox-Snell pseudo-$R^2$), which can be partially attributed to the distribution of the performance data: most participants achieved very high scores, resulting in a ceiling effect in which lower performance levels occur only~infrequently.

The analysis of the distribution of the VA over time revealed distinct peaks corresponding to training videos and additional peaks during validation, interspersed with periods of reduced attention. This temporal dynamic provides empirical evidence that engagement is not a static global state but rather a dynamic process evolving in response to specific task demands and environmental stimuli~\cite{AbeysingheEtAl2025}. {\color{black} The observed attention dips between peaks may operationally index post-presentation cognitive processing windows during which learners consolidate perceived information, mentally rehearse observed sign sequences, or~prepare for the subsequent segment, though~alternative explanations such as natural fatigue or task-switching costs cannot be ruled out based on behavioral data alone.} This alignment between our empirical observations and established principles of cognitive science strengthens the validity of our temporal analysis framework and supports interpretation of observed patterns as authentic reflections of learning processes rather than measurement~artifacts.

These results validate and extend existing research frameworks on learning analytics and engagement measurement through automatically-derived behavioral indicators extracted from interaction traces~\cite{DuEtAl2025,BetetaEtAl2022}. They confirm that in the specific educational domain of VR-based sign language learning, automatically-derived behavioral indicators provide reliable and predictively significant signals of learning performance. The~null finding for VRF adds an important cautionary note to simplistic self-regulated learning~models.

Practically, these results suggest concrete pathways toward attention-aware learning analytics systems in immersive sign language educational applications. VA and PPVT signals could be computed in real time within educational VR applications, enabling adaptive feedback and targeted instructional interventions. However, implementation of such adaptive systems requires rigorous experimental validation before large-scale deployment, as~overly intrusive feedback could produce counterintuitive~effects. 

Despite these interesting implications, the~study presents some limitations. The~sample comprises 117 European university students without significant prior sign language experience; results may differ substantially in more diverse populations including children, older adults, individuals with cognitive disabilities, or~those with prior sign language knowledge. The~learning material was restricted to 12 basic phrases within a coherent narrative context; tasks of greater scope, higher grammatical complexity, larger sign sets, or~decontextualized content could alter engagement dynamics or learning performance substantially. {\color{black} Additionally, while our behavioral indicators serve as operational proxies for underlying cognitive processes, they do not provide direct access to learners' internal mental states. Future work incorporating complementary measures such as think-aloud protocols or physiological indicators could triangulate these findings and provide richer insights into the cognitive mechanisms underlying the observed behavioral patterns.}

Moreover this work is a correlational study, indeed, while our regression model demonstrates robust covariation between VA, PPVT, and~performance, we cannot {\color{black} determine whether high VA or extended PPVT directly cause improved learning outcomes, or~whether both are consequences of other learner characteristics such as motivation or prior VR knowledge. Future experimental studies that systematically manipulate engagement indicators would be necessary to establish causal relationships and test whether interventions targeting these behavioral indicators can effectively enhance learning outcomes.}

Finally, the~study focuses solely on immediate learning within a single session; it is still unclear whether the engagement effects observed persist over longer periods, or~how well brief engagement patterns predict performance in longer-term learning~protocols.

\section{Conclusions and Future~Works}
\label{sec:Conclusions}

This paper shows that objective behavioral traces collected in a VR sign language learning application can predict learning outcomes and reveal time-varying engagement dynamics, with~VA and PPVT emerging as the most informative indicators and VRF providing limited predictive value in this setting.  
In addition, aggregated VA profiles highlight structured engagement patterns across training and validation, reinforcing the value of temporal process analysis alongside outcome~prediction.  

These findings support a practical path toward attention-aware learning analytics in immersive sign language education, where engagement signals can be used not only for post-hoc evaluation but also for designing adaptive support. A~promising direction is to move from prediction to causally grounded design by validating whether real-time, attention-contingent feedback and targeted review policies measurably improve learning efficiency and~retention. 

In the future, we plan to extend our dataset; indeed, the~participants were university students who generally performed well. Future studies involving more heterogeneous populations or more challenging tasks might yield more moderate goodness-of-fit values as the variance in performance increases. Other future work should incorporate richer temporal and sequence-based modeling of interactions (beyond counts) and~assess generalization across different languages, sign sets, and~longer-term learning~protocols.

\vspace{6pt}




\authorcontributions{{Conceptualization},  J.M.A.-L., M.B.-F., D.T., D.U. and E.Y.-B.; Methodology, J.M.A.-L., M.B.-F., D.T. and E.Y.-B.; Resources, J.M.A.-L., D.T. and E.Y.-B.; Data curation, J.M.A.-L., M.B.-F. and E.Y.-B.; Writing---original draft preparation, J.M.A.-L., M.B.-F., D.T. and E.Y.-B.; Writing---review and editing, J.M.A.-L., M.B.-F., D.T., D.U. and E.Y.-B.; Supervision, J.M.A.-L. and E.Y.-B. All authors have read and agreed to the published version of the manuscript.}

\funding{This work was conducted with the support of the project ISENSE: Innovative Supporting
sErvices for uNiversity Students with dEafness funded
by the European Union within the Erasmus+ programme
(call 2022), Key Action 2: KA220 Cooperation Partnerships
for higher education. AGREEMENT n. 2022-1-IT02-KA220-
HED-000089554. Additionally, these results are part of
UNITE: University Network for Inclusive and digiTal Education, project funded by the European Union within the
Erasmus+ programme (call 2023), Key Action 2: KA220
Cooperation Partnerships for higher education (2023-1-IT02-
KA220-HED-0001621181).}

\institutionalreview{The research adhered strictly to ethical
regulations, including Spain’s Organic Law 3/2018 on the Protection of Personal Data and the principles outlined in the
Declaration of Helsinki. Ethical approval was obtained from
the university’s institutional ethics committee (CEIH-25-54) on 11 July~2025.}

\informedconsent{Informed consent was obtained from all subjects involved in the~study.}

\dataavailability{The raw data supporting the conclusions of this article will be made available by the authors on request.}

\acknowledgments{José Manuel Alcalde-Llergo enrolled in the National PhD in Artificial Intelligence, XXXVIII cycle, course on Health and Life Sciences, organized by Università Campus Bio-Medico di Roma. He is also pursuing his doctorate with co-supervision at the Universidad de Córdoba (Spain), enrolled in its PhD program in Computation, Energy and~Plasmas.}

\conflictsofinterest{The authors declare no conflicts of~interest.} 



\abbreviations{Abbreviations}{
The following abbreviations are used in this manuscript: 
\\

\noindent
\begin{tabular}{@{}ll}
Abbreviation 
 & Definition \\
AI & Artificial Intelligence \\
AR & Augmented Reality \\
DHH & Deaf and Hard-of-Hearing \\
ER & Extended Reality \\
GLM & Generalized Linear Model \\
HMD & Head-Mounted Display \\
ISENSE & Innovative Supporting sErvices for uNiversity Students with dEafness \\
PPVT & Post-Playback Viewing Time \\
RQ1 & Research Question 1 \\
RQ2 & Research Question 2 \\
VA & Visual Attention \\
VR & Virtual Reality \\
VRF & Video Replay Frequency \\
\end{tabular}
}

\appendixtitles{no} 

\begin{adjustwidth}{-\extralength}{0cm}

\reftitle{References}

\PublishersNote{}
\end{adjustwidth}
\end{document}